\documentclass[prb,amsmath,amssymb,superscriptaddress,twocolumn]{revtex4}
\usepackage{float}
\usepackage{amsmath}

\usepackage{amssymb}
\usepackage{amsfonts}
\usepackage{euscript}
\usepackage{enumerate}
\usepackage{hhline}
\usepackage{pslatex}
\usepackage{tabularx}
\usepackage[usenames,dvipsnames]{xcolor}

\usepackage{graphicx}

\usepackage{dcolumn}
\usepackage{bm}
\usepackage[sort&compress]{natbib}

\makeatletter
\renewcommand{\@biblabel}[1]{#1. }
\renewcommand{\@dotsep}{500}
\renewcommand{\@pnumwidth}{0em}
\renewcommand{\l@figure}[2]{
\@dottedtocline{1}{1.5em}{2em}{Figure #1}{}\vspace{15pt}}

\usepackage[normalem]{ulem}

\begin{document}
\title{On-chip optical parametric oscillation into the visible: \\ generating red, orange, yellow, and green from a near-infrared pump}

\author{Xiyuan Lu}\email{xiyuan.lu@nist.gov}
\affiliation{Microsystems and Nanotechnology Division, Physical Measurement Laboratory, National Institute of Standards and Technology, Gaithersburg, MD 20899, USA}
\affiliation{Institute for Research in Electronics and Applied Physics and Maryland NanoCenter, University of Maryland,
College Park, MD 20742, USA}
\author{Gregory Moille}
\affiliation{Microsystems and Nanotechnology Division, Physical Measurement Laboratory, National Institute of Standards and Technology, Gaithersburg, MD 20899, USA}
\affiliation{Joint Quantum Institute, NIST/University of Maryland,
College Park, MD 20742, USA}
\author{Ashutosh Rao}
\affiliation{Microsystems and Nanotechnology Division, Physical Measurement Laboratory, National Institute of Standards and Technology, Gaithersburg, MD 20899, USA}
\affiliation{Department of Chemistry and Biochemistry, University of Maryland,
College Park, MD 20742, USA}
\author{Daron A. Westly}
\affiliation{Microsystems and Nanotechnology Division, Physical Measurement Laboratory, National Institute of Standards and Technology, Gaithersburg, MD 20899, USA}
\author{Kartik Srinivasan} \email{kartik.srinivasan@nist.gov}
\affiliation{Microsystems and Nanotechnology Division, Physical Measurement Laboratory, National Institute of Standards and Technology, Gaithersburg, MD 20899, USA}
\affiliation{Joint Quantum Institute, NIST/University of Maryland, College Park, MD 20742, USA}
\date{\today}

\begin{abstract}
     \noindent {Optical parametric oscillation (OPO) in a microresonator is promising as an efficient and scalable approach to on-chip coherent visible light generation. However, so far only red light at $<$ 420~THz (near the edge of the visible band) has been reported. In this work, we demonstrate on-chip OPO covering $>$130~THz of the visible spectrum, including red, orange, yellow, and green wavelengths. In particular, using a pump laser that is scanned 5~THz in the near-infrared from 386~THz to 391~THz, the signal is tuned from the near-infrared at 395~THz to the visible at 528~THz, while the idler is tuned from the near-infrared at 378~THz to the infrared at 254~THz. The widest signal-idler separation we demonstrate of 274~THz corresponds to more than an octave span and is the widest demonstrated for a nanophotonic OPO to date. Our work is a clear demonstration of how nonlinear nanophotonics can transform light from readily accessible compact near-infrared lasers to targeted visible wavelengths of interest, which is crucial for field-level deployment of spectroscopy and metrology systems.}
\end{abstract}

\maketitle
\begin{figure*}[t!]
\centering\includegraphics[width=1.0\linewidth]{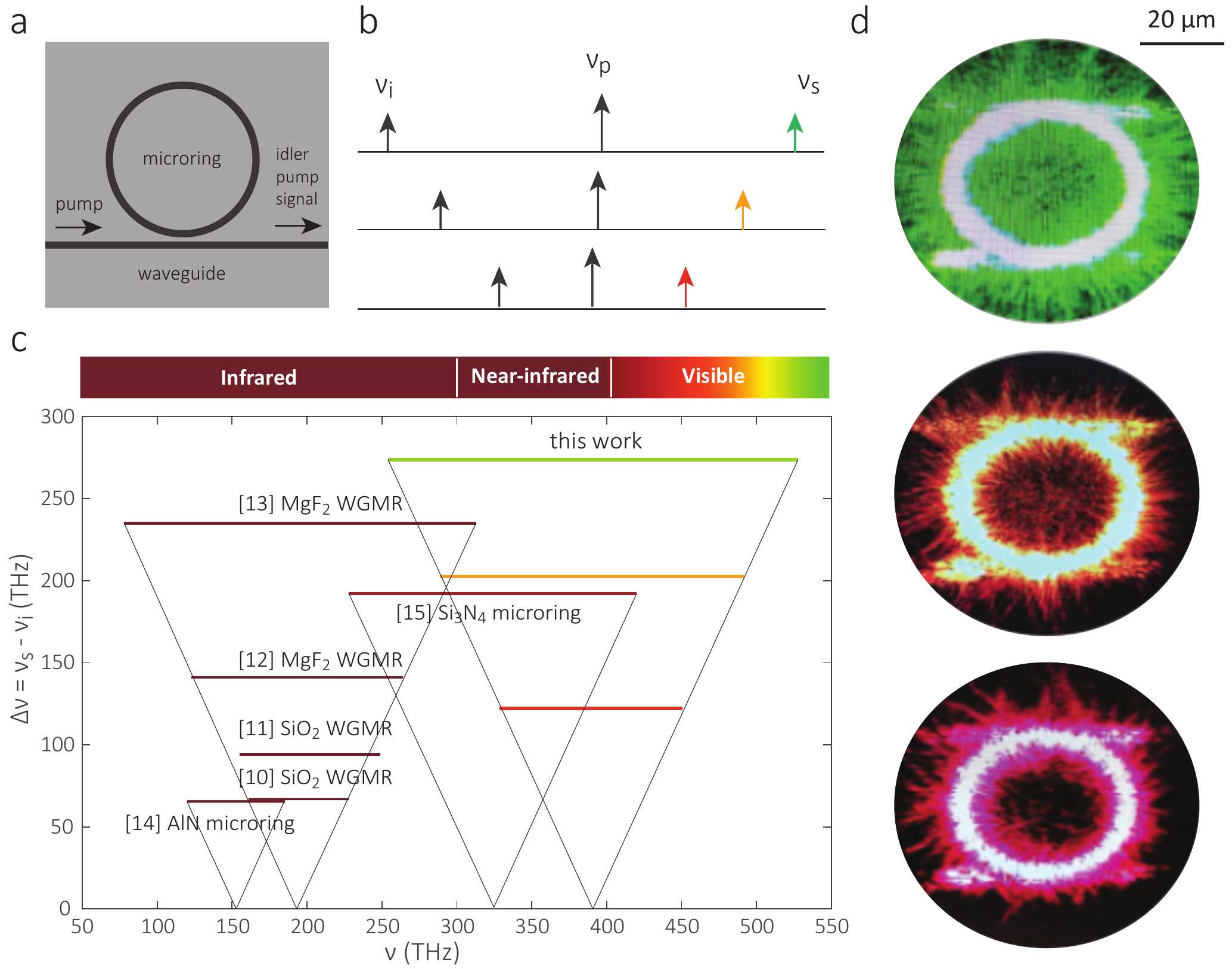}
\caption{{\bf Optical parametric oscillation (OPO) transforms a near-infrared pump into visible wavelength light.} \textbf{a,~b,} Schematic showing the microring device and the OPO to generate signal and idler in the visible and infrared, respectively, using a pump in the near-infrared. When the pump shifts by only a few THz, the OPO outputs (signal in the visible and idler in the infrared) shift by $\approx$~100~THz. This advantage makes OPO unique for light generation over broad spectral bands. \textbf{c}, Comparison of widely-separated OPO in miniaturized/on-chip photonic devices. Most of these works are pumped by infrared lasers either at $\approx$ 200~THz (1.5~$\mu$m) for MgF$_2$ and SiO$_2$ whispering gallery mode resonators (WGMRs) or at $\approx$ 150~THz (2~$\mu$m) in an AlN microring, where the representative data with the widest span (y axis) are shown by the brown lines. The OPO output signal and idler have all been limited to being in the infrared and near-infrared. A recent OPO work in Si$_3$N$_4$ uses a near-infrared pump at $\approx$ 330~THz (910~nm) to generate red light at the edge of the visible at $\approx$~420~THz (714~nm), as shown by the dark red line. The contribution of this work is to go deeper into the visible, covering from red (bottom) to green (top), using only a few THz of pump tuning around $\approx$ 385~THz (780~nm). In particular, the top line (green) shows the case where the OPO is octave-spanning, with idler and signal bridging infrared and green wavelengths. 
\textbf{d,} Top view microscope images of devices exhibiting OPO, showing the scattering of generated visible light in the red, orange, and green. These images are taken without ambient illumination, using a camera that is insensitive to the pump and generated idler wavelengths. The device is a Si$_3$N$_4$ microring with an outer radius of 25~$\mu$m. The 20~$\mu$m scale bar in the top right corner applies to all three images. Two waveguides are designed for each microring, as evident from the scattered light in the images, but only the bottom waveguide is used to effectively couple pump, signal, and idler light, as depicted in (a).}
\label{Fig1}
\end{figure*}

\noindent On-chip generation of coherent visible light is important for scalable manufacturing and field-level deployment of many applications in spectroscopy, metrology, and quantum optics. For example, many wavelength references~\cite{Hollberg_2005} and optical clocks~\cite{Ludlow2015} are based on visible lasers stabilized to atomic systems. Moreover, many quantum systems suitable for local storage and manipulation of quantum information, including trapped ions, atoms, and spins in crystals, have optical transitions that require visible pump lasers for operation~\cite{Simon2010}. A direct approach is to develop on-chip lasers based on III-V semiconductors, e.g., indium gallium arsenide lasers~\cite{Sun2016}, but spectral coverage is typically limited by the available gain media. Achieving spectral coverage via direct optical transitions over a wide range of wavelengths is challenging, and integrating such a laser into a mature photonic integrated circuit platform is also nontrivial.

Optical parametric oscillation (OPO) using a second-order nonlinearity ($\chi^{(2)}$) or a third-order nonlinearity ($\chi^{(3)}$) is a unique process for creating light with wide spectral coverage using only a single pump laser~\cite{Boyd2008,Agrawal2007}. In particular, over the past few decades, table-top $\chi^{(2)}$ OPO has been extensively studied for visible light generation~\cite{Ebrahimzadeh1989}, and has been a major workhorse for laboratory purposes. Such table-top OPO provides a versatile source that offers high-power tunable laser light, but is also bulky, expensive, and lacks the miniaturization and scalability for field-level deployment in many applications.

Historically, OPO has utilized the $\chi^{(2)}$ nonlinearity more often than the $\chi^{(3)}$ nonlinearity because of its larger nonlinear response in bulk materials~\cite{Boyd2008}. However, nanophotonic technologies enable strong enhancement of light intensities in time and space, and make both $\chi^{(2)}$ and $\chi^{(3)}$ processes power-efficient. In fact, $\chi^{(3)}$ processes are even comparable to $\chi^{(2)}$ processes in certain cases~\cite{Lu2019B}. In recent years, $\chi^{(2)}$ nanophotonic OPO has been reported in an aluminum nitride microcavity~\cite{Tang2019_Optica}, where telecom signal and idler beams are generated by a near-infrared pump laser. However, $\chi^{(2)}$ nanophotonic OPO faces a major challenge for visible light generation, that is, the energy conservation criterion requires a UV pump laser for the signal to cover the visible spectral range ($\omega_\text{p}=\omega_\text{s}+\omega_\text{i}$, where p, s, i represent pump, signal, idler, respectively). This requirement, along with the large overall spectral separation between pump, signal, and idler, makes $\chi^{(2)}$ nanophotonic OPO very challenging to design and integrate for visible light generation. In contrast, $\chi^{(3)}$ nanophotonic OPO is naturally suited for generating visible light. For such an OPO to generate visible (signal) and infrared (idler) light, the energy conservation criterion (2$\omega_\text{p}=\omega_\text{s}+\omega_\text{i}$) only requires near-infrared lasers, which are commercially available in compact forms that are ready for chip-integration.

The development of a $\chi^{(3)}$ OPO device that covers the visible spectrum from 405~THz (red) to 790~THz (violet) with a near-infrared pump source is a challenging goal. There have been several pioneering works demonstrating widely-separated OPO in the infrared~\cite{Fujii2017, Sayson2017,Fujii2019,Sayson2019,Tang2019}; however, in that wavelength range the constituent materials are much less dispersive than in the visible. Recently, OPO based on silicon nitride (Si$_3$N$_4$) microrings generated red light at 420~THz (714~nm) by a 325~THz (920~nm) pump~\cite{Lu2019C}. Though milliwatt-level threshold power and the ability to suppress competing processes was shown, the OPO output only reached the long wavelength edge of the visible spectrum.

In this work, we demonstrate OPO in a Si$_3$N$_4$  microring (Fig.~\ref{Fig1}(a)) that addresses $\approx$~34~\% (130~THz) of the visible spectrum, including red, orange, yellow, and green colors, through a small change in the pump laser frequency (Fig.~\ref{Fig1}(b)). Our approach enables octave-spanning OPO to be observed, with a 527.8~THz (568.4~nm) signal in the green and a 254.1~THz (1181~nm) idler in the infrared. The corresponding span of 273.7~THz is the widest on-chip OPO reported so far \textendash~it is $>$~40~THz larger than the previous record~\cite{Sayson2019} set for an infrared OPO (Fig.~\ref{Fig1}(c)) \textendash~despite the aforementioned increase in material dispersion at visible wavelengths. We further show that through power tuning, the visible output signal can be finely tuned similar to devices with larger footprints. Our work represents a major advance in using nonlinear nanophotonics to access desired wavelengths in the visible spectrum (Fig.~\ref{Fig1}(d)), and may have numerous applications in spectroscopy, metrology, and quantum science.
\begin{figure}[htbp]
\centering\includegraphics[width=1.0\linewidth]{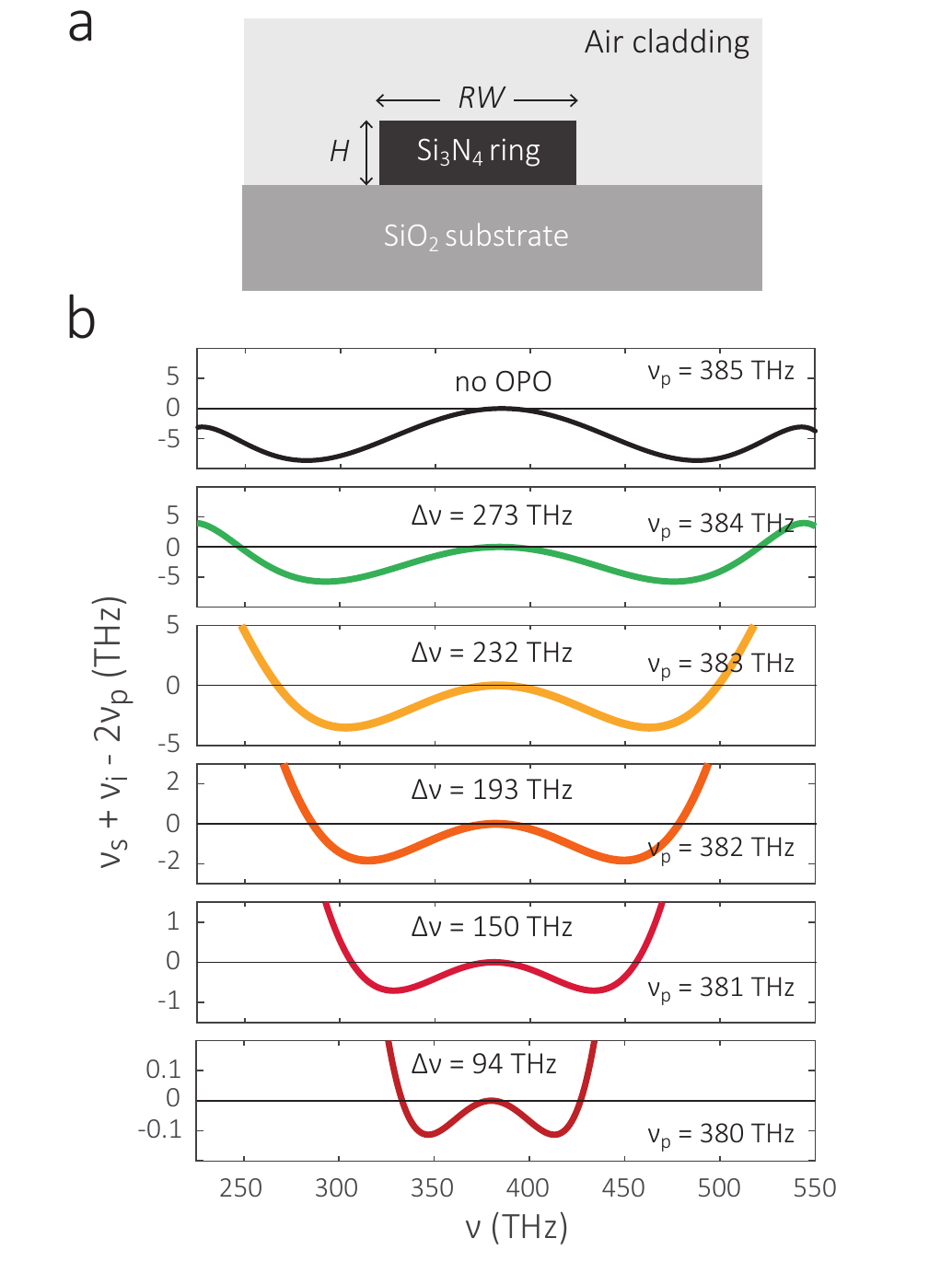}
\caption{{\bf Dispersion design for visible light generation.} \textbf{a,} Cross-section view of the device shows two geometric parameters, ring width ($RW$) and film thickness/height ($H$), that are critical for dispersion engineering. \textbf{b,} Simulation shows the frequency matching profiles that support the OPO spectral separations depicted in Fig.~\ref{Fig1}(c). Frequency mismatch (y axis) is plotted for phase matched mode sets ($m_\text{s} + m_\text{i} - 2 m_\text{p} = 0$), where the pump is shifted from 380~THz (bottom) to 385~THz (top) by a step of 1~THz. In the top panel, when the pump is at 385~THz (black), the device dispersion is too large to support any OPO. When the pump is decreased to 384~THz (green), the device supports octave-spanning OPO with idler and signal at 247.5~THz and 520.5~THz, respectively. A further decrease in the pump frequency leads to a smaller frequency span with different colors for the visible wavelength signal (e.g., yellow, orange, and red). The device has a radius of 25~$\mu$m, a thickness of 500 nm, and a radius of 825~nm.}
\label{Fig2}
\end{figure}

\bigskip \noindent \textbf{Results}

\noindent \textbf{Design and simulation} \label{SectA}
In recent years, there have been several studies emphasizing the Si$_3$N$_4$ nonlinear photonic platform~\cite{Moss2013} as being especially suitable for wide-band nonlinear optics. These studies include demonstrations of octave-spanning microresonator frequency combs~\cite{Okawachi2011, Li2017, Karpov2018}, frequency conversion for quantum and classical applications~\cite{Li2016, Lu2019B}, and entangled photon-pair generation for quantum communication~\cite{Lu2019C}.

The physical process to support widely-separated OPO is cavity-enhanced degenerate four-wave mixing. To achieve such a nonlinear optical process, both the momentum and energy for the interacting cavity modes have to be conserved~\cite{Boyd2008, Agrawal2007}. In particular, when modes from the same mode family (e.g., the fundamental transverse-electric mode TE1 in this work) are used, momentum conservation reduces to a simple equation, $m_\text{s}+m_\text{i}-2m_\text{p} = 0$, where the subscripts $s,i,p$ denote signal, idler, and pump, respectively, and $m$ is the azimuthal mode number. Energy conservation requires frequency matching, that is, the frequency mismatch ($\nu_\text{s}+\nu_\text{i}-2\nu_\text{p}$) needs to be within the cavity linewidths ($\nu_k/Q_k$, where $k=s,i,p$ and $Q_k$ is the loaded quality factor for the $k$ mode). As four-wave mixing can occur across multiple sets of modes simultaneously, another important factor in device design is that all other signal and idler mode sets do not simultaneously realize frequency and phase matching~\cite{Lu2019C}. For example, close-to-pump OPO has been a major competitive process for widely-separated OPO when the device exhibits anomalous dispersion around the pump~\cite{Vahala2004}.

We simulate the azimuthal numbers ($m$) and corresponding frequencies ($\nu_\text{m}$) of TE1 modes using the finite-element method for a device with ring radius ($RR$) of 25 $\mu$m, thickness ($H$) of 500~nm, and ring width ($RW$) of 825~nm (Fig.~\ref{Fig2}(a)). For each configuration that satisfies phase matching, i.e., $m_\text{s}+m_\text{i}-2m_\text{p} = 0$, the frequency mismatch is plotted in Fig.~\ref{Fig2}(b). When the pump laser is at $\nu_\text{p}=$385~THz, no modes are frequency and phase matched for OPO. When the pump laser is situated below 385~THz, OPO starts to appear and its span decreases as the pump frequency decreases. For the widest OPO, the span is $\approx$ 273~THz, with the signal predicted to be in the green.

Importantly, with this design all of the aforementioned pump frequencies are in the normal dispersion regime, that is, $\nu_\text{s}+\nu_\text{i}-2\nu_\text{p} <0$ when $\nu_\text{s}$ and $\nu_\text{i}$ are close to $\nu_\text{p}$. Such dispersion prohibits close-to-pump OPO because any Kerr shift further decreases this frequency mismatch (more negative for OPO) when pump power is injected into the cavity. Therefore, we expect that our visible-infrared OPO devices should be free from competition due to close-to-pump OPO.

\begin{figure}[htbp]
\centering\includegraphics[width=1.0\linewidth]{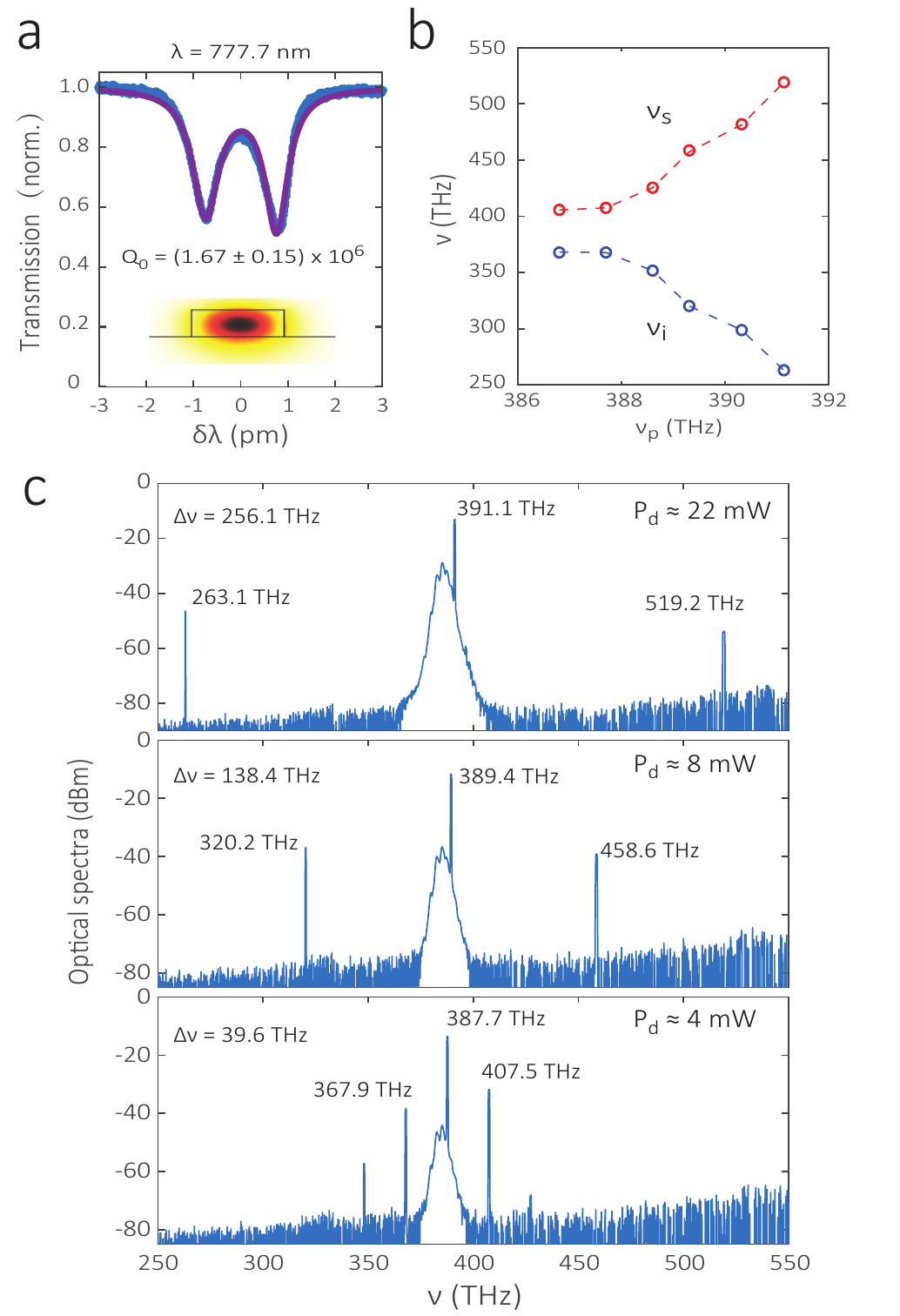}
\caption{{\bf Widely-separated OPO in a nominal device.} \textbf{a,} Normalized transmission of a pump mode (fundamental transverse-electric mode, whose electric field profile is in the inset) shows an intrinsic Q of (1.67~$\pm$~0.15)$\times$10$^6$, where the uncertainty is a one standard deviation value obtained from the fitting of the doublet resonance. \textbf{b,} When the pump frequency shifts by $\approx$ 5~THz, the signal (red) and idler (blue) frequencies shift by 123~THz and 113~THz, respectively. The ratio of the signal/idler spectral tuning to that of pump is $\approx$ 26:1. \textbf{c,} Three optical spectra are shown with pump frequency of 391.1~THz, 389.4~THz, and 387.7~THz, from top to bottom, respectively. In general, the required pump power dropped into the cavity is larger when OPO span is larger. The spectra are clean without any noise or competitive processes. In the bottom spectrum, a second pair of OPO sidebands emerges and is much weaker ($\approx$ 20~dB) than the primary pair, and is commonly observed for OPO that is close to pump. The device has a radius of 25~$\mu$m, a thickness of 500~nm, and a nominal ring width of 826~nm.}
\label{Fig3}
\end{figure}

\bigskip \noindent \textbf{Widely-separated OPO into the visible} \label{SectB}
The devices are fabricated with nominal parameters of fixed thickness and ring radius ($H$ = 500~nm, $RR$ = 25~$\mu$m) while varying ring widths ($RW$ = 820~nm to 830~nm), and are characterized as a function of $\nu_{\text{p}}$ near the simulated near-infrared frequencies around 385~THz. For example, the characterization of widely-separated OPO in a device with nominal $RW$ of 826~nm is shown in Fig.~\ref{Fig3}.

The output OPO spectra are recorded by an optical spectrum analyzer (OSA) as the pump is tuned over modes that support widely-separated OPO with normal dispersion (387~THz to 391~THz), as shown in Fig.~\ref{Fig3}(b). The pump mode has intrinsic $Q$ of 1$\times$10$^6$ to 2$\times$10$^6$ and loaded $Q$~$\approx$~1$\times$10$^6$. For example, we show in Fig.~\ref{Fig3}(a) a pump mode at 385.8~THz (777.7~nm), which is a fundamental transverse-electric mode (TE1), as shown in the inset. When the pump frequency is $>$~391~THz, no widely-separated OPO is observed. When the pump frequency is at 391~THz, the OPO device has the widest span of 256.1~THz (the top panel of Fig.~\ref{Fig3}(c)). The signal is at 519.2~THz (577.8~nm), which is yellow in color. The idler is at 254.1~THz (1140~nm) in the infrared. A further decrease in the pump frequency shifts the signal to red at 458.6 THz (654.1~nm) and then towards the near-infrared at 407.5~THz (736.2~nm), and shifts the idler from within the infrared at 320.2~THz (936.9~nm) to the near-infrared at 367.9~THz (815.4~nm), as shown in the bottom two panels of Fig.~\ref{Fig3}(c).

With this device, a mere $\approx$ 5~THz pump shift leads to a 123~THz shift of the signal and a 113~THz shift of the idler. Such large amplification in the output tuning range relative to the pump tuning range comes from the large dispersion of the nanophotonic resonator in the targeted frequency matching bands, and is particularly useful when wide spectral coverage is needed.

\begin{figure*}[t!]
\centering\includegraphics[width=1.0\linewidth]{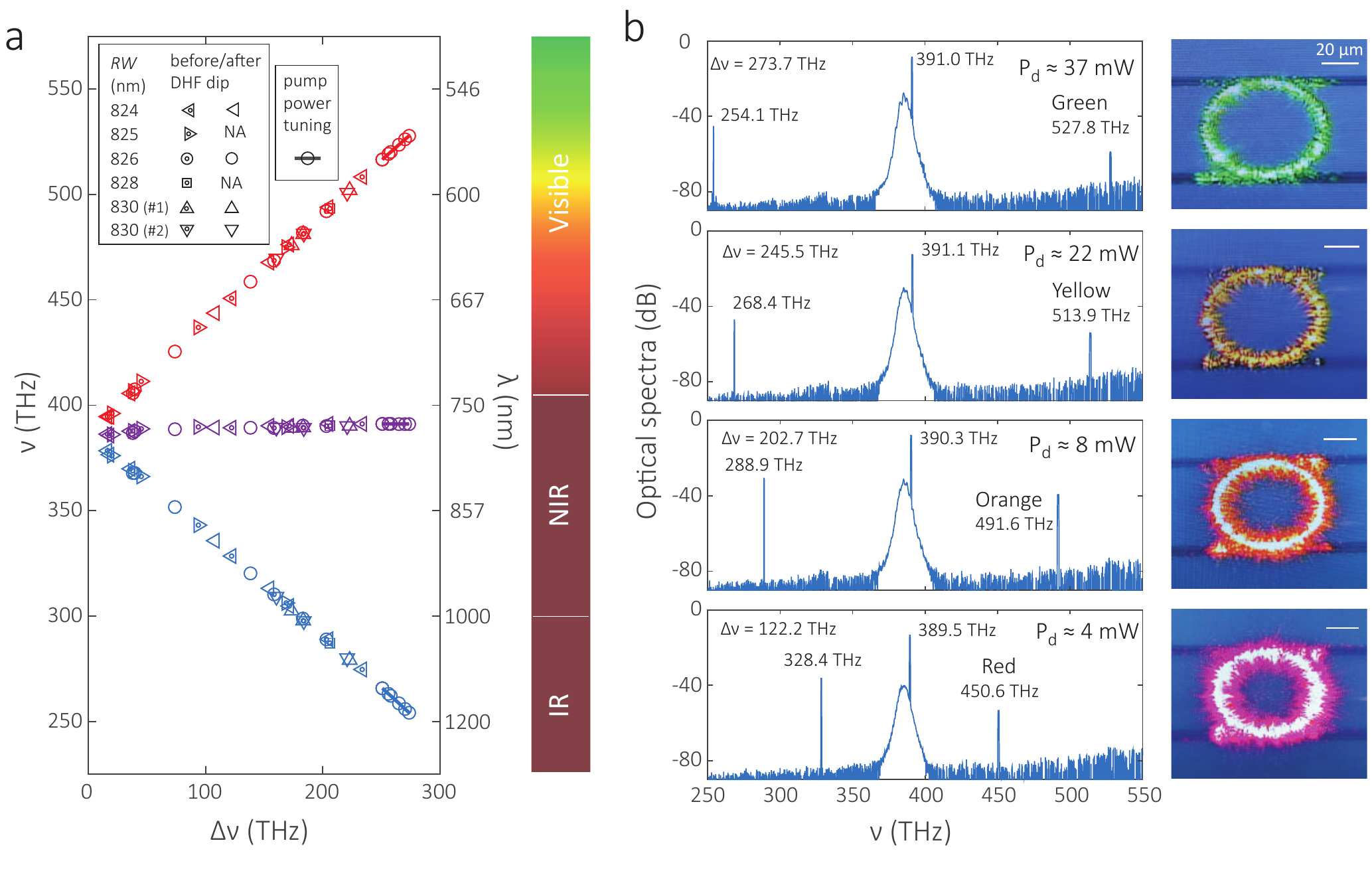}
\caption{{\bf OPO output light from red to green.} \textbf{a}, A collection of signal (red), pump (violet), and idler (blue) frequencies plotted versus the OPO span ($\Delta \nu$). Pump laser is scanned from 386~THz to 391~THz, and its span in use is only 5~THz. The OPO signal frequencies span $>$~130~THz, corresponding to a 26 times amplification. This figure shows data from six devices with ring widths from 824~nm to 830~nm before and after DHF dip. The DHF dip is dilute (100:1) hydrofluoric acid etching that isotropically removes $<$~1~nm of Si$_3$N$_4$. The overall frequency span covers red, orange, yellow, and green signal wavelengths. In particular, OPO frequencies with $\Delta \nu$ from 256.1~THz to 273.7~THz are accessed by pump power tuning (Fig.~\ref{Fig5}). \textbf{b}, Representative OPO spectra for signal output light in the green at 527.8~THz (568.4 nm), yellow at 513.9 THz (583.7 nm), orange at 491.6 THz (610.3 nm), and red at 450.6 THz (665.8 nm) from top to bottom, respectively. Optical microscope images (with illumination) show the corresponding colors that are generated. The scale bars represent 20~$\mu$m in these four images.}
\label{Fig4}
\end{figure*}

The required pump power ($P_\text{d}$), specified in Fig.~\ref{Fig3}(c), has a dependence on the OPO span. $P_\text{d} = P(1-T)$ represents the pump power that is coupled into the microring, where $P$ is the input pump power in the waveguide and $T$ is the normalized cavity transmission of the pump laser mode. For example, in the top panel, an OPO spanning 256.1 THz requires a pump power of $P$ $\approx$ 42~mW in the waveguide, of which $P_\text{d}$ $\approx$~22~mW is dropped into microring to excite the OPO above threshold. In the middle and bottom panel, where the OPO span is 54~\% and 15~\% of the widest span, the required dropped power is 36~\% and 18~\% of that in the top panel. Such power dependence may come from lower $Q$ of the cavity resonances at higher frequencies.

\bigskip \noindent \textbf{Light generation from red to green} \label{SectC}
The wavelengths at which light is generated are very sensitive to the device geometry because of the large amplification of the frequency span relative to the pump detuning from its nominal position. As shown in the previous section, even a change in pump mode of one free spectral range (FSR $\approx$~0.9~THz) leads to a $>$~20~THz change in the frequency of the visible signal. Similarly, a small change in device dimension, even on the order of 1~nm, can lead to a different color of light generated in the visible band. This dispersion sensitivity to geometry requires approaches to realize fine spectral tuning for these nanophotonic OPO devices to be useful in practice, but also provides coarse spectral coverage using only a few devices with parameters close to the nominal design. For example, in Fig.~\ref{Fig4}(a), we collect the pump, signal, and idler frequencies from different pumping modes in six devices with ring widths from 824~nm to 830~nm. We can see that within a 10 THz pump scan in the near-infrared (violet symbols), the OPO outputs of these devices taken together cover $>$~270 THz, spanning from the infrared and near-infrared to the visible. This spectral coverage, although still discrete, provides better coverage than that of the single device shown in Fig.~\ref{Fig3}(b). In particular, this set of devices covers $>$~130 THz of the visible band, including red, orange, yellow, and green colors (see the right colorbar in Fig.~\ref{Fig4}(a)).

We present four optical spectra showing the generation of green, yellow, orange, and red light in Fig.~\ref{Fig4}(b). In each case, the OPO spectrum shows no noise or competitive nonlinear processes, and its color is confirmed by optical microscope images when the device is in operation, showing red, orange, yellow, and green light generated in the microring. In particular, in a device with $RW$ = 826~nm, when the pump mode is at 391.0~THz (the top panel), we observe the widest OPO with a span of 273.7~THz, which is very close to the simulated value (273~THz). The pump mode frequency is $\approx$ 7 THz larger than the simulation (384~THz), which is likely due to a combination of uncertainty in the refractive index model chosen for Si$_3$N$_4$ and in the fidelity of the fabricated device dimensions relative to design. This octave-spanning OPO has a 527.8~THz (568.4~nm) signal and a 254.1~THz (1181~nm) idler.

\bigskip \noindent \textbf{Fine spectral coverage with power tuning} \label{SectD}
The ability to realize continuous spectral coverage would complement the above demonstration of broad spectral coverage using the widely-separated OPO process. The broad  coverage is made possible by the dispersion properties of the resonator, which results in an amplification of the signal tuning range relative to the pump laser tuning range of $\approx$ 26:1 (Fig.~\ref{Fig4}(a)). Of course, the resonant nature of the device is such that the pump laser is not tuned continuously, but instead in jumps from across different pump modes separated by FSR $\approx$ 0.9~THz. As a result, the output spectral coverage, though broad in overall extent, is sampled with both signal and idler frequencies in jumps of multiple FSRs, as shown in Fig.~\ref{Fig4}(a). In comparison, larger devices used in other OPO works in the infrared \cite{Sayson2017,Fujii2017, Fujii2019,Sayson2019} obtain a finer step in output frequency. In this section, we show that we can achieve fine spectral coverage of the signal and idler in the visible and infrared bands, respectively, by shifting one pump mode continuously through pump power tuning.

\begin{figure}[t!]
\centering\includegraphics[width=1.0\linewidth]{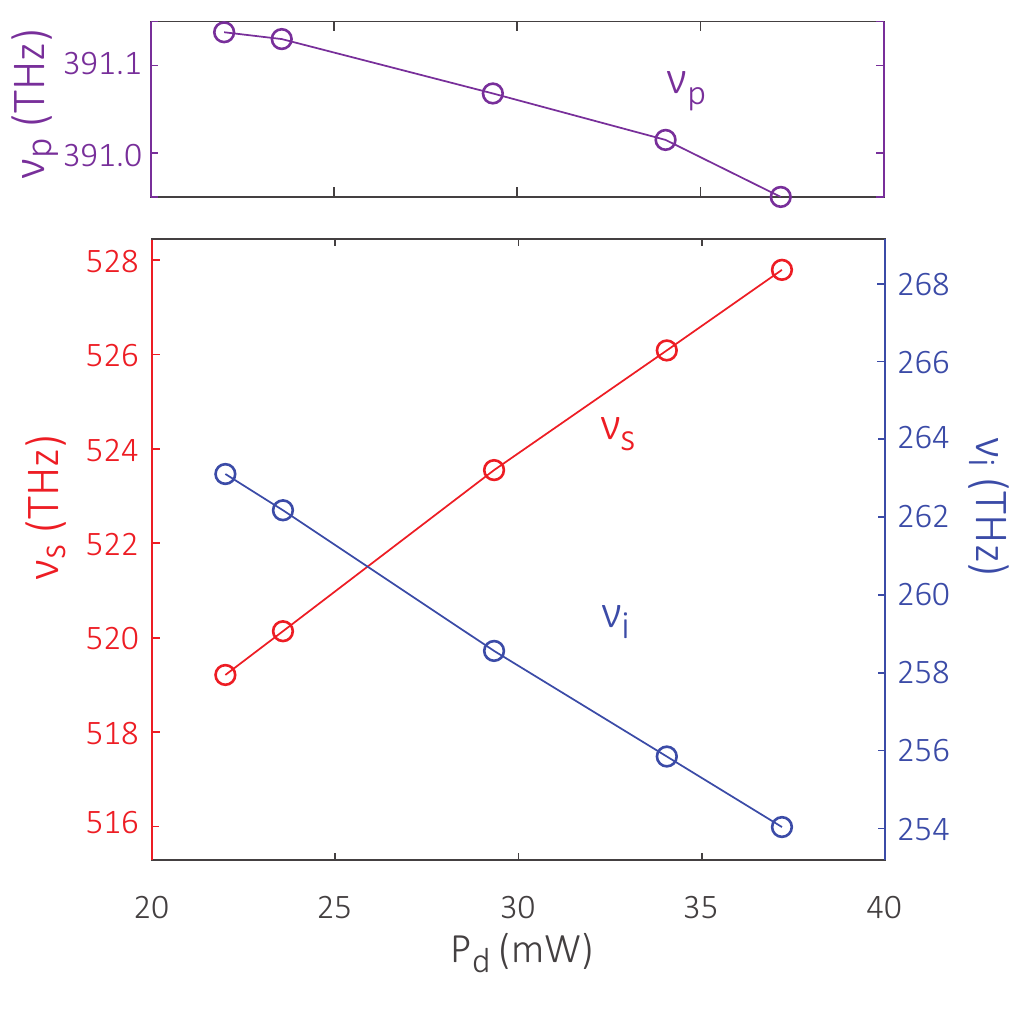}
\caption{{\bf Tuning of OPO frequencies with pump power.} When the dropped pump power ($P_\text{d}=P(1-T)$, where $P$ is the laser power in the waveguide and $T$ is the normalized cavity transmission of the pump mode that can be adjusted by laser-cavity detuning) increases from 22 mW to 37 mW, the pump frequency ($\nu_\text{p}$) is adjusted to follow the thermo-optical shift of the pump mode.}
\label{Fig5}
\end{figure}

An example of this fine coverage is shown in Fig.~\ref{Fig5}. This device generates green light as shown in the top panels of Fig.~\ref{Fig3}(c) and Fig.~\ref{Fig4}(b). The OPO frequencies depend on the pump power dropped into the microring ($P_\text{d}$), which results in a thermo-optical shift that linearly depends on $P_\text{d}$. We plot the pump (violet), signal (red), and idler (blue) frequencies versus $P_\text{d}$ in Fig.~\ref{Fig5}. When $P_\text{d}$ is 22~mW, the idler is at 263.1~THz and signal is at 519.2~THz (the top panel of Fig.~\ref{Fig3}(c)). When $P_\text{d}$ increases to 37~mW, the idler frequency decreases to 254.1~THz and signal frequency increases to 527.8~THz (the top panel of Fig.~\ref{Fig4}(b)). The visible frequency shifts by 8.6 THz, while the pump frequency shifts by $<$ 0.2~THz. Therefore, such thermo-optic tuning method can be used to improve the spectral coverage in the visible band, bringing the multiple-FSR jumps down to about 1 FSR. In practice, further improvement towards a continuous coverage may require using larger devices with smaller FSRs~\cite{Fujii2017,Sayson2017,Fujii2019,Sayson2019,Tang2019}, at the cost of increased threshold power, and cascading or coupling several devices with slightly different dimensions, at the cost of simplicity of usage.

\bigskip \noindent \textbf{Discussion}
\noindent In summary, we propose and demonstrate, for the first time, nanophotonic OPO devices whose output frequencies (including both the signal and idler) cover a range from 527.8 THz to 254.1 THz, which encompasses the green, yellow, orange, and red parts of the visible spectrum, as well as the near-infrared and a portion of the infrared spectrum. Our OPO magnifies frequency span of the near-infrared pump source by $\approx$~25 times in the generated frequency ranges for both infrared and visible light. Devices whose dispersion supports the widest separation in signal and idler frequencies exhibit octave-spanning OPO at $<$~30~mW threshold power. We further show a tuning method to achieve fine spectral coverage. Our work is a major advance in the realizing coherent on-chip sources of  visible light. One future challenge is in reaching higher frequencies in the visible (e.g., cyan and blue light), which will require further investigation into appropriate dispersion design, resonator-waveguide coupling, and the potential impact of increased intrinsic losses in Si$_3$N$_4$  at those colors.

\noindent \textbf{Acknowledgements} This work is supported by the DARPA DODOS and NIST-on-a-chip programs. X.L. acknowledges support under the Cooperative Research Agreement between the University of Maryland and NIST-PML, Award no. 70NANB10H193.

\bibliography{OPO780}

\end{document}